\documentclass[fleqn,10pt]{wlscirep}
\usepackage[utf8]{inputenc}
\usepackage[T1]{fontenc}
\usepackage{authblk}
\usepackage[utf8]{inputenc}
\usepackage{graphicx}
\graphicspath{{images/}}
\usepackage{times} % Use {S times fonts
\usepackage{amsmath}
\usepackage{amsfonts}
\usepackage{ dsfont }
\usepackage{ gensymb }
\usepackage{url}
\usepackage{physics}
\usepackage{cite}
\usepackage[linesnumbered,ruled,vlined]{algorithm2e}
\usepackage{algpseudocode}
\usepackage{soul}
\usepackage{color}
\usepackage{mathtools}

\def\bs{\begin{subequations}}
\def\es{\end{subequations}}
\def\aa{\begin{align}}
\def\ab{\end{align}}
\def\ba{\begin{eqnarray}}
\def\ea{\end{eqnarray}}
\def\be{\begin{equation}}
\def\ee{\end{equation}}

\newcommand{\la}{\langle}
\newcommand{\ra}{\rangle}

\def\ben{\begin{enumerate}}
\def\een{\end{enumerate}}
\def\bs{\bigskip}

\def\ss{\smallskip}

\title{Optimizing target nodes selection for the control energy of directed complex networks}

\author[1]{Hong Chen}
\author[1,*]{Ee Hou Yong}
\affil[1]{Division of Physics and Applied Physics, School of Physical and Mathematical Sciences, Nanyang Technological
University, Singapore, 637371, Singapore. }
\affil[*]{ Correspondence and requests for materials should be addressed to E.H.Y.
(email: eehou@ntu.edu.sg)
}
\date{}

\begin{document}

\flushbottom

\begin{abstract}
The energy needed in controlling a complex network is a problem of practical importance. Recent works have focused on the reduction of control energy either via strategic placement of driver nodes, or by decreasing the cardinality of nodes to be controlled. However, optimizing control energy with respect to target nodes selection has yet been considered. In this work, we propose an iterative method based on Stiefel manifold optimization of selectable target node matrix to reduce control energy. We derive the matrix derivative gradient needed for the search algorithm in a general way, and search for target nodes which result in reduced control energy, assuming that driver nodes placement is fixed. Our findings reveal that the control energy is optimal when the path distances from driver nodes to target nodes are minimized. We corroborate our algorithm with extensive simulations on elementary network topologies, random and scale-free networks, as well as various real networks. The simulation results show that the control energy found using our algorithm outperforms heuristic selection strategies for choosing target nodes by a few orders of magnitude. Our work may be applicable to opinion networks, where one is interested in identifying the optimal group of individuals
that the driver nodes can influence.
\end{abstract}

\maketitle

\section{Introduction}

Complex networks have been extensively studied in recent decades owing to its modeling utility towards social systems \cite{albert2002statistical}, biological systems \cite{jeong2000large}, Internet \cite{albert2000error}, and man-made technological systems \cite{ruths2014control}. Usually, these networks are modelled as coupled system of ordinary differential equations. The state vector elements are represented as nodes or vertices in graphs and the coupling interaction between state vector elements are represented as links or edges in graphs. The state vectors of such a coupled ordinary differential equation serve to represent a myriad of quantities, depending on the complex dynamical system being considered at hand. For example, they can represent the probability of a person being infected in a social network system, or they can represent the expression level of a gene in a regulatory network\cite{barzel2013universality}.  The motivation to study and understand these complex systems can be traced to our desire to obtain control over them \cite{liu2011controllability}. In this case, control refers to exerting influence on the networked system via external control signals to steer the state vector of the networked system from its arbitrary initial, to a predefined goal state vector in finite time $[t_0,t_f]$ \cite{liu2011controllability}. It follows then, that if the nodes of a network could be steered towards the predefined goal state vector in finite time, the network is deemed controllable.

Achieving control over a complex network with as few control signals as possible is desirable. In 2011, Liu et al proved that the unmatched nodes from maximum matching algorithm \cite{hopcroft1973n} of a bipartite representation of a complex network needed to receive external influence to ensure network structural controllability \cite{lin1974structural,liu2011controllability}. The ${N_d}$ number of unmatched nodes in need of control signals are thus termed the minimum driver node set, or simply the driver nodes. Soon after, Sun and Motter explored the numeric success rate of network controllability when numerically computing the controllability Gramian matrix when using energy optimal control signal \cite{rugh1996linear,kirk2012optimal} to steer the network \cite{sun2013controllability}. They found that for a complex network with more than a handful of nodes, using the minimum driver node set is computationally insufficient as the computation of the controllability Gramian will become ill-conditioned or nearly singular. Instead, beyond using the ${N_d}$ number of minimum driver nodes, additional control signals are needed to ensure numeric success when computing the controllability Gramian.

Since the problem of minimum driver node set to guarantee controllability for an  arbitrary sized complex network was solved \cite{liu2011controllability}, several other prominent research works have followed \cite{cowan2012nodal,klickstein2017locally,yan2012controlling}. Notably, the investigation into the energy cost required by a control signal has been a subject of investigation by several groups \cite{ding2017key,li2016optimal,li2015minimum,yan2012controlling,yan2015spectrum,klickstein2017energy}. In these studies, the energy cost is defined as a measure of proportionate effort exerted by the control signal over the considered time \cite{rugh1996linear,kirk2012optimal}. It was found that if the number of control signals is small, the energy cost demanded of each of the signal could be prohibitively high \cite{yan2015spectrum}. In fact, the energy cost is reduced exponentially as the number of control signal increases \cite{yan2015spectrum}. Thus, attaching additional control signals onto a networked system beyond the minimum driver node set is one way to achieve network control with reduced energy cost.

The way in which the additional control signals are attached can also result in reduced control energy. Lindmark and Altafini considered the eigenvalues of the network and proposed strategies for selecting the placement of additional control signals to minimize control energy cost \cite{lindmark2018minimum}. Chen et al analyzed stems, obtained from minimum driver node set \cite{liu2011controllability}, and calculated all possible direct shortest paths from driver nodes to non-driver nodes to obtain the Longest Control Chain (LCC), which is the longest path from the obtained all possible shortest direct paths \cite{chen2016energy}. They found that by adding additional control signals in such a way that the length of the LCC is minimized, control energy can be reduced significantly. Li et al proposed an algorithm using matrix derivative projected gradient descent to iteratively search for the energy optimal placement of control signals \cite{li2015minimum}. This optimization model was later simplified by Ding et al \cite{ding2017key} and Li et al \cite{li2016optimal}. In 2018, Li et al proposed an improved and generalized approach based on previous works to again obtain the energy-optimal placement of control signals \cite{li2018optimization}. The problem of reducing control energy by strategic placement of additional control signals has been extensively researched. However, in all of these works, complete control was considered, where the control signals steer the full node set towards the predefined goal state vector.

While full control may be necessary in some types of engineered systems \cite{stankovic2012deadline}, controlling just a subset of nodes (typically termed target control or targeted control) may be more sensible in large complex dynamical systems. In 2014, Gao et al proposed an alternate k-walk theory and a greedy algorithm to obtain the minimum number of driver nodes to control just a subset of the full node set in a complex network \cite{gao2014target}. In 2015, Iudice et al presented a geometric framework to find the driver node set, limited by practical constraints, to reach as many target nodes as possible \cite{iudice2015structural}. In 2017, Liu et al considered the controllability of the giant connected component in a directed complex network \cite{liu2017controllability}. With regards to energetic considerations, Klickstein et al showed that the energy cost scales exponentially with the cardinality of the target node set \cite{klickstein2017energy}. Gao et al proposed an algorithm to obtain the placement of control signals to optimize the energy cost when controlling just a subset of nodes in directed complex networks \cite{gao2018towards}. Furthermore, recently, the controllability Gramian of lattice graphs was studied, which turns out to be useful when the number of target nodes is small \cite{klickstein2020controllability}. While target control has been researched extensively, how to optimize control energy with respect to the selection of target nodes in a complex network is still an open question. 

Thus, different from all previous works, we wish to ask a slightly different question: How can we pick the target node set in such a way that the control energy is minimized? To that end, we will be employing a cost function optimization model based on projected gradient descent, similar to earlier works \cite{ding2017key,li2016optimal,li2015minimum,gao2018towards,li2018optimization}. The main difference between our work and the existing literature is the variable matrix being optimized: While previous works have focused on optimizing the choice of input signal (matrix $B$), we optimize the choice of target node set (matrix $C$). Furthermore, previous derivations of the index notation gradient information uses the I-Chain rule, which can be difficult to understand. Here, we derive the index notation gradient information using the standard chain rule and product rule in a general way, which is simpler to follow.

In this paper, we will be examining how the choice of target node set could be optimized to minimize the energy cost function. Using the formulated energy cost function, we derive the matrix derivative of the energy cost function with respect to matrix $C$, which is the energy cost function gradient information. With the gradient information obtained, we perform an iterative search using the trace-constraint-based projected gradient method (TPGM) proposed in ref. \cite{li2018optimization} to obtain the node set which are energetically favorable. We then compare the energy cost that we would have gotten from choosing to target control nodes using a heuristic selection scheme such as random selection, and node degree-based selection. Our simulation results show that the solution obtained from TPGM reduces the energy cost by a few orders of magnitude.

\section{Problem formulation}

In standard complex network controllability literature, we are interested in studying $N$ coupled system of equations whose dynamics are linear and time invariant (LTI):
\begin{equation}
\begin{aligned}
&\dot{\bf x}(t) = A{\bf x}(t) + B{\bf u}(t), \\
&{\bf y}(t) = C{\bf x}(t),
\end{aligned}
\end{equation}
\noindent where ${\bf x}(t) {\in} \mathbb{R}^{N\times1} $ is the time-varying state vector, ${\bf y}(t) {\in} \mathbb{R}^{P\times1}$ ($P \leq N$) is the the subset of time-varying output state vector that we want to target control. ${\bf u}(t) {\in} \mathbb{R}^{M \times 1}$ ($M \leq N$) is a time-varying external control signal which we use to drive the network. The time invariant $A \in \mathbb{R}^{N \times N} $ matrix represents the network topology, where for directed complex network we have nonzero matrix element $\{a_{ij}\}$ when there is a directed link from node j to node i, and zero element if no link exists from node j to node i. Since many complex systems tend to enjoy passive stability, our modeling also allows for self-links, where $\{a_{ii}\}$ is a negative real number \cite{cowan2012nodal}. The time invariant matrix $B \in \mathbb{R}^{N \times M} $  reflects the coupling between nodes and $M$ number of external control signals ${\bf u(t)}$, where matrix element $\{b_{ij}\}=1$ if node i receives a time-varying control signal from control input j, and zero element otherwise. Finally, we have the the control matrix $C \in \mathbb{R}^{P \times N}$ which tells us the choice of the $P$ number of subset of nodes that we wish to target control, where matrix element $\{c_{ij}\}=1$ if node j is the $i$-th node out of all possible $P$ nodes that we want to target control and zero otherwise. We note that matrix $B$ and matrix $C$ are column and row linear independent, where if $\{b_{ij}\}=1$ ($\{c_{ij}\}=1$), then no other nonzero entries may exist for column $j$ and row $i$ in matrix $B$ ($C$). We require this linear independence to reflect our modeling choice for one-to-one connection between control signals and nodes, and for one-to-one connection to individually target control each of the $P$ number of nodes \cite{klickstein2017energy}.

We are interested in the energy cost needed to drive the states of the network using the control signal {\bf u}(t), which is defined to be the cost function \cite{rugh1996linear}
\be
J=\int_{t_0}^{t_f}{\bf u}^{T}(t){\bf u}(t)dt.
\ee 
This is the energy cost function that we want to optimize, with respect to the choice of target nodes (matrix $C$). Similar to previous works which considered trace constraint projected matrix gradient cost function optimization\cite{ding2017key,li2016optimal,li2015minimum,gao2018towards,li2018optimization}, here we assume that the initial state vector ${\bf x}_0 \sim \mathcal{N} (0,1)$. The energy cost is 
\be \label{energycost_general}
\mathcal{E}(t_f,B,C)=\mathbb{E}[\int_{0}^{t_f}{\bf u}^{T}(t){\bf u}(t)dt],
\ee
where $t_f$ is the total control time and initial time is set to zero throughout the rest of the paper, $t_0 =0$. The energy cost is a function of the control trajectory, i.e. the state space pathway between ${\bf x}_0$ and ${\bf y}_f$.  It should be noted that (\ref{energycost_general}) is the expected energy cost over all realizations of initial state vector ${\bf x}_0$, picked under standard normal distribution. In general, the cost function $\mathcal{E}(t_f,B,C)$ is dependent on the matrix $B$ and the final time $t_f$. However, in this paper, we are focusing on optimizing the cost function with respect to matrix $C$ while keeping input matrix $B$ and final time $t_f$ fixed. Thus, for clarity, we will drop these two dependencies and write the cost function only as a function of either $\mathcal{E}(C)$ or $\mathcal{E}(C^T)$ throughout the paper. 

Consequently, the energy-optimal control signal, derived from optimal control theory \cite{kirk2012optimal,klickstein2017energy} can be calculated as
\begin{equation} \label{optimal control signal}
\begin{aligned}
&{\bf u}^{*}(t) = B^{T}e^{A^{T}(t_{f}-t)}C^{T}(CWC^{T})^{-1}({\bf y}_f -
Ce^{A(t_f-t_0)}{\bf x}_0).
\end{aligned}
\end{equation}
The $N \times N$ controllability Gramian matrix 
\be
W=\int_{t_0}^{t_f} e^{A(t_f-t)}BB^{T}e^{A^{T}(t_f-t)} dt
\ee
is an important quantity in control theory, well-known to be real, symmetric, and semi-positive definite. In practice, we can use the controllability Gramian matrix to verify that the choice of matrix $B$ is ensuring network (target) controllability by checking that the (output $(CWC^{T})$) controllability Gramian matrix is invertible \cite{rugh1996linear,klickstein2017energy}.

We optimize the cost function as follows:
\begin{equation} \label{optimise}
\text{min} \quad \mathcal{E}(C^T) \quad \text{subject to} \quad tr(CC^T) = P,
\end{equation}
where we have constrained the solution produced by the gradient descent iterative algorithm to stay on the manifold $\tr(CC^T)=P$.

\section{Results}

The energy cost function can be solved by substituting the energy-optimal control signal ${\bf u}^{*}(t)$ into the cost function $\mathcal{E}(C^T)=\mathbb{E}[\int_{0}^{t_f}{\bf u}^{T}(t){\bf u}(t)dt]$:
\begin{equation} \label{energy cost}
\begin{aligned}
& \mathcal{E}(C^T)=tr((CWC^{T})^{-1}{\bf y}_f{\bf y}_f^{T})+tr(C^{T}(CWC^{T})^{-1}Ce^{At_f}e^{A^{T}t_f}).
\end{aligned}
\end{equation}
By varying Eq.~\ref{energy cost} with respect to $C^T$, we find that
\begin{equation} \label{energy gradient}
\begin{aligned}
\frac{\partial \mathcal{E}(C^T_k)}{\partial C^T}=&-2WC^T(CWC^T)^{-1}{\bf y}_f{\bf y}_f^T(CWC^T)^{-1}\\&-2WC^T(CWC^T)^{-1}Ce^{At_f}e^{A^Tt_f}C^T(CWC^T)^{-1} \\
&+2e^{At_f}e^{A^Tt_f}C^T(CWC^T)^{-1}.
\end{aligned}
\end{equation}
The full derivation can be found in the Supplementary Information. Using (\ref{energy cost}) and (\ref{energy gradient}) and applying it to the TPGM algorithm, we solve for the energy-optimal target node set to target control and obtain the optimal solution, ${ C^{*}}$. In the simulations, we have set the control time to be $t_f=2$, and the desired final output state vector is ${\bf y}_f = [1,1,...,1]^T$.

\subsection{Algorithm} \label{Algorithm}
To optimize the control energy by varying the selection of target nodes, we modified the trace-constraint-based projected gradient method (TPGM) formulated by Li et al \cite{li2018optimization}, to focus on $C^T$, the target nodes choice. %\subsubsection{Preliminaries}
%Let $\mathbb{O} _1^{N \times P} :=\{C^T \in \mathbb{R}^{N \times P}:tr(CC^T)=P\}$, and
Let $\tilde{C}^T$ be the basis of the target control matrix $C^T$. We can obtain $\tilde{C}^T$ by performing Gram Schmidt orthogonalization on matrix $C^T$ \cite{gram-schmidt}. Define the projection operator  $\mathcal{T}_{\tilde{C}^T}=(I_{N}-\tilde{C}^T\tilde{C})$, where $I_N$ is the $N \times N$ identity matrix. The operator $\mathcal{T}_{\tilde{C}^T}$ projects any arbitrary matrix onto the space that is perpendicular to the manifold $tr(CC^T)=P$. 
% $\mathcal{M} := \{X: X=\tilde{C}^TY,\quad Y \in \mathbb{R}^{P \times P} \}$. We note that with each iteration update of the solution, the matrix is updated by a small increment in the direction that is perpendicular to its previous iteration solution
For two arbitrary matrices $A$ and $B$ which have the same dimension, an angle between them can be defined as:
\begin{equation} \label{angle}
\theta = \arccos \bigg(\frac{tr(A^T B)}{\|A \|_F \|B \|_F}\bigg),
\end{equation}
where $ 0\leq \theta \leq \pi$, $\|.\|_F$ denotes Frobenius norm, and we note that the matrices are perpendicular if $\theta = \pi/2 = 90 \degree $. The TPGM optimization is given in Algorithm~\ref{Pseudocode}.

\begin{algorithm}
\caption{Pseudocode for finding energy-optimal target control matrix $C^*$ using the trace-constraint-based projected gradient method (TPGM). }
\SetAlgoLined
\label{Pseudocode}
\DontPrintSemicolon
Initialize $C^{T}_{k=0}$ as a random $N\times P$ matrix

\While{$\cos(\theta_k) > \xi$}{
Compute $\frac{\partial \mathcal{E}(C^T_k)}{\partial C^T}= \nabla \mathcal{E}(C^T_k)$ \;
Update $\hat{C}^T_{k+1} =C^T_k - \eta\cdot(I_N-\tilde{C}^T_k\tilde{C}_k)\nabla\mathcal{E}(C^T_k)$\;
Normalize $C^T_{k+1}=\sqrt{\frac{P}{tr(\hat{C}_{k+1}\hat{C}^T_{k+1})}} \cdot \hat{C}^T_{k+1}$\; 
Compute $\cos (\theta_k) = \Bigg(\frac{tr\Big([\nabla \mathcal{E}(C^T_k) ]^T \mathcal{T}_{\tilde{C}^T_k} \nabla \mathcal{E}(C^T_k) \Big)}{\|\nabla \mathcal{E}(C^T_k) \|_F \cdot \| \mathcal{T}_{\tilde{C}^T_k} \nabla \mathcal{E}(C^T_k)\|_F}\Bigg)$ \;
Update $k=k+1$}
\end{algorithm}

In TPGM step $1$, by random initialization of matrix $C^T_0$, we mean that we start with a $N \times P$ matrix of zeros and randomly set matrix element $[C^T]_{ij}=1$, where node $i$ is chosen to be the $j$-th target node. Once node $i$ is chosen, we maintain row and column linear independence, and thus do not allow for the selection of the same node $i$ to be target controlled. In the numerical experiments that we tried, while starting from a pure random dense matrix is possible when system size is small or the network topology is sparse, it is computationally inefficient. Furthermore, when starting from a pure random dense matrix, we are not able to obtain a sensible sparse optimal matrix at the end, when converting from dense optimal solution $C^{*}$ to sparse binary optimal solution (see section \ref{select from real to discrete}). 

In TPGM step $4$, the numerical value of the learning rate or step size $\eta$ is chosen empirically. When $\eta$ is too large, convergence is not guaranteed and the algorithm will fail. While convergence is guaranteed when $\eta$ is sufficiently small, the time taken to complete the iterative search will suffer if $\eta$ is too small. Typically, we choose the learning rate to be between $1$e-$8$ and $1$e-$3$, depending on the fraction $P/N$ to be target controlled as well as the complex network topology. In general, the learning rate $\eta$ can be varied to speed up the iterative process. For example, starting at $\eta = 1$e-$8$, and then changing to $\eta = 1$e-$4$ when enough iterations have been run. As observed during experimentation,  the convergence of $\eta$ scales inversely proportional to the numerator of the $\cos (\theta_k) $: $tr\Big([\nabla \mathcal{E}(C^T_k) ]^T \mathcal{T}_{\tilde{C}^T_k} \nabla \mathcal{E}(C^T_k) \Big)$. In TPGM step $4$, $\hat{C}^T_{k+1}$ is a non-normalized quantity, while in step $5$, the updated solution is constrained onto the manifold surface $tr(CC^T)=P$, obtaining the normalized $C^T_{k+1}$. 

In TPGM step $6$, the angle $\theta_k$ between $k$-th step gradient matrix $\nabla \mathcal{E}(C^T_k)$, and the projected gradient matrix, $ \mathcal{T}_{\tilde{C}^T_k} \nabla \mathcal{E}(C^T_k)$ is calculated to check for convergence. If $\cos (\theta_k)$ approaches zero, or equivalently, when $\theta_k$ approaches $\pi/2 = 90\degree$, the algorithm is deemed to have converged. Numerically, the while loop terminating condition, $\xi$, refers to a small positive quantity, for example, $\xi=1$e-$2$. For some networks where the algorithm is unable to converge towards $\xi=1$e-$2$, we may relax this condition to $\xi=1$e-$1$.

TPGM will iteratively update the initial proposed solution $C_{0}$ in the direction of quickest decreasing energy cost based on cost function derivative with respect to matrix variable $C$. When the search has finally converged, the obtained optimal target control matrix $C^{*}$ corresponds to reduced control cost of a dense matrix of real numbers, which corresponds to many-to-many connections from output nodes to complex network nodes of varying link strength. Based on the obtained optimal solution $C^{*}$, the challenge is to obtain a one-to-one connection of output nodes and nodes with unity link strength, a sparse binary optimal solution, $C^{*}_{\text{binary}}$, while maintaining the characteristics of reduced control cost.    

\subsection{Selecting binary optimal target node set from $C^{*}$} \label{select from real to discrete}

We propose two methods to find $C^{*}_{\text{binary}}$ from $C^{*}$. The first method is based on suppressing insignificant matrix elements in $C^{*}$, and then evaluating the normalised quantity \textit{importance index} vector, $r$:

\begin{equation} 
\begin{aligned}
&r = \frac{[r_1,r_2,...,r_N]}{max(r_1,r_2,...,r_N)} \\
\end{aligned}
\label{importance index1}
\end{equation}

\noindent where $r_i = \sum_j |[C^{T}]_{ij}|$ is the non-normalised importance score of node $i$ being selected as a target node, and we are taking the row summation of the absolute of the transpose matrix, $C^T$. $|[C^T]_{ij}|$ represents the numerical contribution towards $C^{*}$ for node $i$, target node $j$. Therefore, to find binary optimal solution that remains similar to $C^{*}$, we want to find the nodes with the highest numerical contributions. 

In our experimentation, we found that directly passing $C^{*}$ into (\ref{importance index1}) to find $C^{*}_{\text{binary}}$ usually does not allow us to find the optimal target node set. This is because of the row summation of the absolute of the numerical contribution. For example, if row $a$ has predominantly absolute values of around $0.2$ in all its columns, while row $b$ has an absolute value of around $0.9$ in one of its columns, and mostly negligible numerical values close to zero in all of its other columns, then by equation (\ref{importance index1}), node $a$ will be ranked higher than node $b$. Based on experimentation with elementary topologies, we find that numerical contribution characteristics similar to row $b$ usually correspond to an optimal target node. Thus, we want the row summation to reflect that: we suppress the numerical values in each matrix element of $C^{*}$ if they fall below a certain value. (For a more detailed discussion, see Supplementary Information.)

We compute the suppressed optimal matrix by setting $[C^{*}]_{ij}=0$ iff $[C^{*}]_{ij}<= d \; \sigma$, where $\sigma$ is computed from the standard deviation of the absolute of all matrix elements $[C^{*}]_{ij}$, and $d=\{0.0,0.1,0.2,...,1.1,1.2,...,3.0\}$. For each suppressed matrix, we check that the rank of the matrix is $P$, before considering it as a viable candidate solution. If the suppressed matrix has rank $P$, then we pass the suppressed matrix into (\ref{importance index1}), and pick $P$ nodes with the highest importance index. The case when $k=0$ is similar to the proposed importance index vector formulation in refs \cite{li2015minimum,li2016optimal,li2018enabling} for finding optimal driver nodes. 

The second method is based on selecting the matrix elements with the largest absolute numerical values. The process is as follows: First, begin with optimal matrix transpose, $|[C^{*}]^T|$. Then, search for the absolute largest matrix element in each column, and order the columns in descending order, keeping track of the associate row indices. Start with a $N \times P$ zero matrix, $C^T_{\text{binary}}$, and set the matrix element for position $(i,j)$ to be one, starting from the columns with the largest absolute matrix elements $|[C^*]_{ij}^T|$. At each step, check that for assigning position $(i,j)$ to be one, no other nonzero element exists along row $i$ or along column $j$. If there exists another nonzero matrix element along the row $i$ or column $j$, then do not assign position $(i,j)$ to be one, and record down column $j$ which did not get filled. If rank$(C^T_{\text{binary}})=P$, then stop the process; otherwise, repeat the process described for unfilled columns $j$'s for the case of second largest absolute matrix elements, third largest,...., until rank$(C_{\text{binary}})=P$. 

At the end, we pass all the obtained binary target node set candidate solutions into the objective cost function, equation (\ref{energy cost}), and pick the target node set which yields the lowest energy cost, denoting it $[C^{*}_{\text{binary }}]^t$, where $t=\{1,2,...,10\}$ represents each independent iterative search. For the simulation results showing $C^{*}_{\text{binary}}$, we choose the best solution out of all $[C^{*}_{\text{binary }}]^t$.

\subsection{Numerical experiments on elementary network topologies}

To understand the arrangement of energy optimal target node set, we perform the TPGM iterative search on elementary network topologies \cite{liu2011controllability,li2018enabling}. They are: directed stem, circle, and dilation. A stem requires just a control signal, placed at the root, to become mathematically (but not necessarily numerically) controllable \cite{liu2017controllability,wang2017physical,sun2013controllability}. A circle is controllable with just one control signal attached to any of the nodes in the circle. A dilation requires a minimum of two driver nodes to become controllable. We select the driver nodes to be placed in such a way that their path distances are evenly spaced, which corresponds to the most energy efficient set up as the longest path distance from driver to non-driver nodes is minimized \cite{chen2016energy}. 
%We maintain the driver nodes to be fixed and we allow, for controlling $66.7 \%$ of the nodes in the elementary topologies, the target node set to be variable. 
The set of driver nodes is assumed to be fixed ($B$ fixed), and the target node set, consisting of $66.7 \%$ of all nodes, is our variable, e.g. find $C$.

\begin{figure}[h!] 
\begin{center}
\includegraphics[scale=0.05]{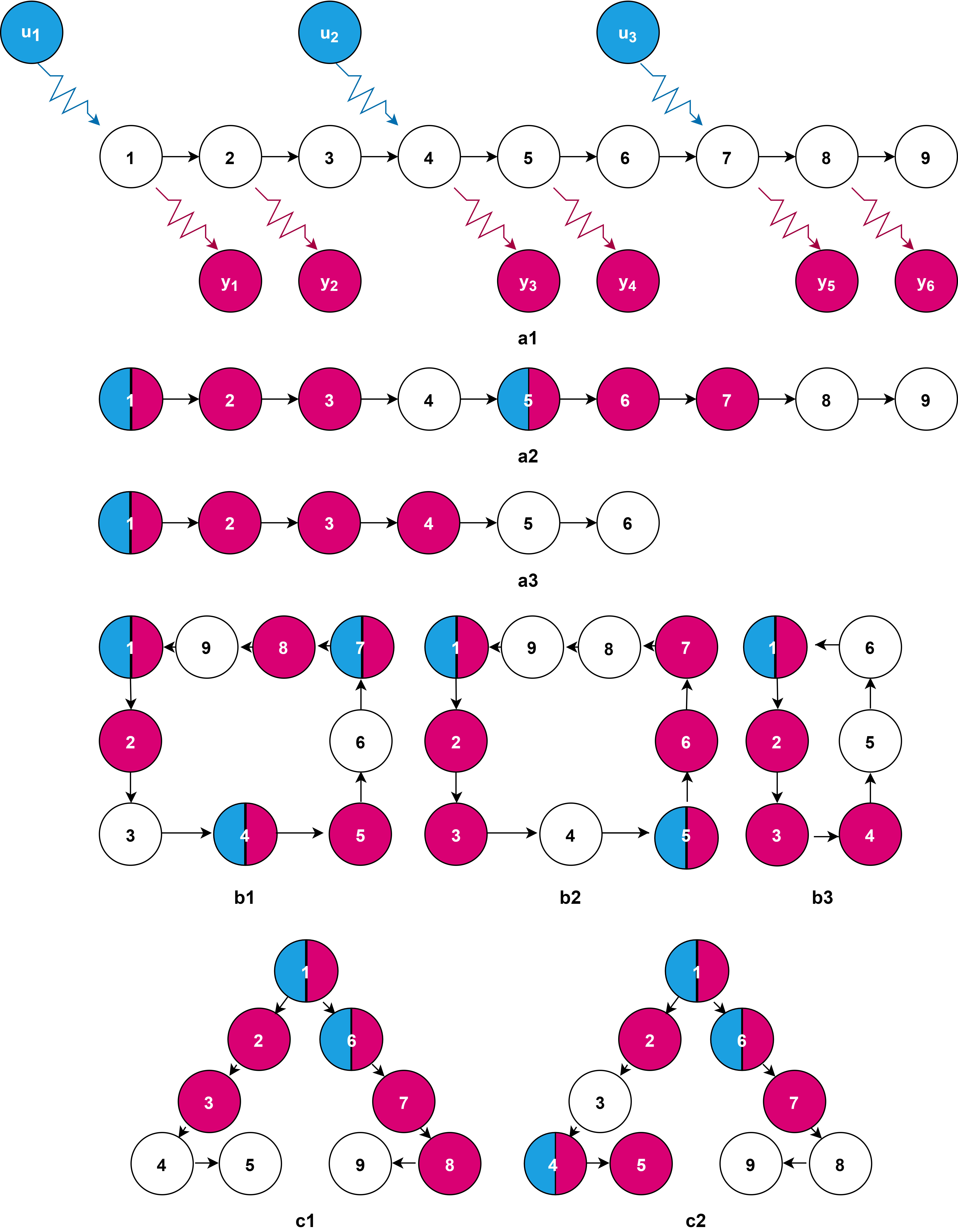}
\end{center}
\caption{Energy optimal target nodes configurations when driver nodes are fixed. {\bf (a1)} depicts a $N=9$ directed stem network, with $M=3$ driver nodes depicted by the cyan (grey in grayscale) control signals, and $P=6$ optimal target nodes depicted by the magenta (dark grey) output nodes. {\bf (a2-a3)} represent $\{N=9,M=2,P=6\}$ and $\{N=6,M=1,P=4\}$ directed stem network respectively. {\bf (b1-b3)} represent circle topology with $\{N=9,M=3,P=6\}$, $\{N=9,M=2,P=6\}$, and $\{N=6,M=1,P=4\}$ respectively. {\bf (c1-c2)} represent dilation with $\{N=9,M=2,P=6\}$ and $\{N=9,M=3,P=6\}$ respectively. } 
\label{fig:elementary topologies}
\end{figure}

For each of the $8$ elementary topology labelled {\bf (a1)} to {\bf (c2)} as shown in Fig.~\ref{fig:elementary topologies}, we repeat the numerical experiment independently for $1000$ times in order to find the optimal target node set $C^*_{\text{binary}}$, given that driver nodes position are fixed. {\bf (a1)} is a $N=9$ directed stem network, with $M=3$ driver nodes, and $P=6$ optimal target nodes. {\bf (a2-a3)} represent $\{N=9,M=2,P=6\}$ and $\{N=6,M=1,P=4\}$ directed stem network respectively. {\bf (b1-b3)} represent circle topology with $\{N=9,M=3,P=6\}$, $\{N=9,M=2,P=6\}$, and $\{N=6,M=1,P=4\}$ respectively. {\bf (c1-c2)} represent dilation topology with $\{N=9,M=2,P=6\}$ and $\{N=9,M=3,P=6\}$ respectively. Because the system sizes are small, the global optimal target nodes can be corroborated through brute force computation: {\bf (a1)} - $\{1,2,4,5,7,8\}$, {\bf (a2)} - $\{1,2,3,5,6,7\}$, {\bf (a3)} - $\{1,2,3,4\}$, {\bf (b1)} - $\{1,2,4,5,7,8\}$, {\bf (b2)} - $\{1,2,3,5,6,7\}$, {\bf (b3)} - $\{1,2,3,4\}$, {\bf (c1)} - $\{1,2,3,6,7,8\}$, {\bf (c2)} - $\{1,2,4,5,6,7\}$. Correspondingly, TPGM was able to find the energy optimal target nodes configurations with success rates: {\bf (a1)} - probability $58.9\%$, {\bf (a2)} - $66.0\%$, {\bf (a3)} - $85.9 \%$, {\bf (b1)} - $41.0 \%$, {\bf (b2)} - $40.1 \%$, {\bf (b3)} - $67.4 \%$, {\bf (c1)} - $91.2 \%$, {\bf (c2)} - $69.1 \%$. Evidently, as seen in Fig. \ref{fig:elementary topologies}, the energy optimal target nodes tend to minimize their geodesic \cite{newman2003structure} path distances from the driver nodes. Our findings are consistent with literature, which states that control energy cost increases exponentially with path distance \cite{chen2016energy,klickstein2018control}.

To test the robustness of the algorithm, we repeat the experiments for elementary topologies with fixed driver nodes placement similar to configurations {\bf(a1)}, {\bf(a2)}, {\bf(b1)}, {\bf(b2)}, {\bf(c1)}, and {\bf(c2)} shown in Fig.~\ref{fig:elementary topologies}, but with total target control nodes changed from $P = 6$ to $P=5$. In general, the conclusion that energy-optimal configurations correspond to minimized path distances from driver to target nodes still holds. For the stem network with $3$ driver nodes, the optimal configurations of target nodes are $\{1,2,4,5,7 \}$, $\{1,2,4,7,8 \}$, and $\{1,4,5,7,8 \}$, given in ascending order of energy costs, with the first node set being the true optimal. TPGM was able find the optimal configurations with probabilities $30.8\%$, $15.2 \%$, and $0\%$ respectively. For stem network with $2$ driver nodes, the optimal configurations are $\{1,2,3,5,6 \}$ and $\{1,2,5,6,7 \}$, with the former being the true optimal. TPGM success rates are $15.1\%$ and $55.3\%$ respectively. Thus, while control energy is minimized when target nodes are close to their nearest driver nodes, for stem network, there is a slight reduction in control energy when the target nodes are also closer to node $1$, which is the root node needed to ensure controllability. It should be noted that the energy cost difference between local optimal configurations and global optimal configurations is typically smaller than $10 \%$. For circle topology with $3$ driver nodes, the optimal configurations are $\{1,2,5,7,8\}$, $\{2,4,5,7,8\}$, $\{1,2,4,5,8\}$ and $\{1,2,4,7,8\}$, $\{1,2,4,5,7 \}$, $\{1,4,5,7,8 \}$, with the first three node sets being the true optimal. TPGM was able to find node sets belonging to true optimal energy cost $29.6 \%$ of the time, and optimal configurations with probability $5.7\%$. Circle topology with $2$ driver nodes has optimal configurations $\{1,2,3,5,6 \}$ and $\{1,2,5,6,7\}$, with the former being the true optimal. Correspondingly, the search success rates were $5.5 \%$ and $34.0 \%$ respectively. While a circle is controllable with just one driver node placed anywhere, the difference in energy cost can be accounted for by grouping all output nodes to their nearest driver nodes: $\{1,2,3,4\}$ to node $1$, and $\{5,6,7,8,9\}$ to node $5$. Thus, for energy-optimal control of $5$ target nodes, the first $4$ have to be picked to be close to their nearest driver nodes, and for the fifth node, a slight preference is given to picking target nodes belonging to the group with the lower directed path distance. In all of these, the control energy difference between true optimal and optimal are at most $2\%$. For dilation networks with $2$ driver nodes, the optimal configurations are $\{1,2,6,7,8\}$ and $\{1,2,3,6,7 \}$, with the former being the true optimal. They were respectively found by the algorithm $32.7 \%$ and $59.6 \%$ of the time. The optimal configurations of dilation networks with $3$ driver nodes are $\{1,2,4,6,7\}$, $\{1,2,4,5,6\}$, and $\{1,4,5,6,7\}$, given in ascending order of energy costs. The first node set was found $97.0\%$ of the time and the latter two are never found. The difference in energy cost between the true optimal and optimal is about $10 \%$, and analogous explanation for grouping output nodes to their nearest driver nodes, similar to circle topologies, can be used to explain the slight difference in control energy. 

\subsection{Simulations on a small random network}
\begin{figure}[h!] 
\begin{center}
\includegraphics[scale=0.04]{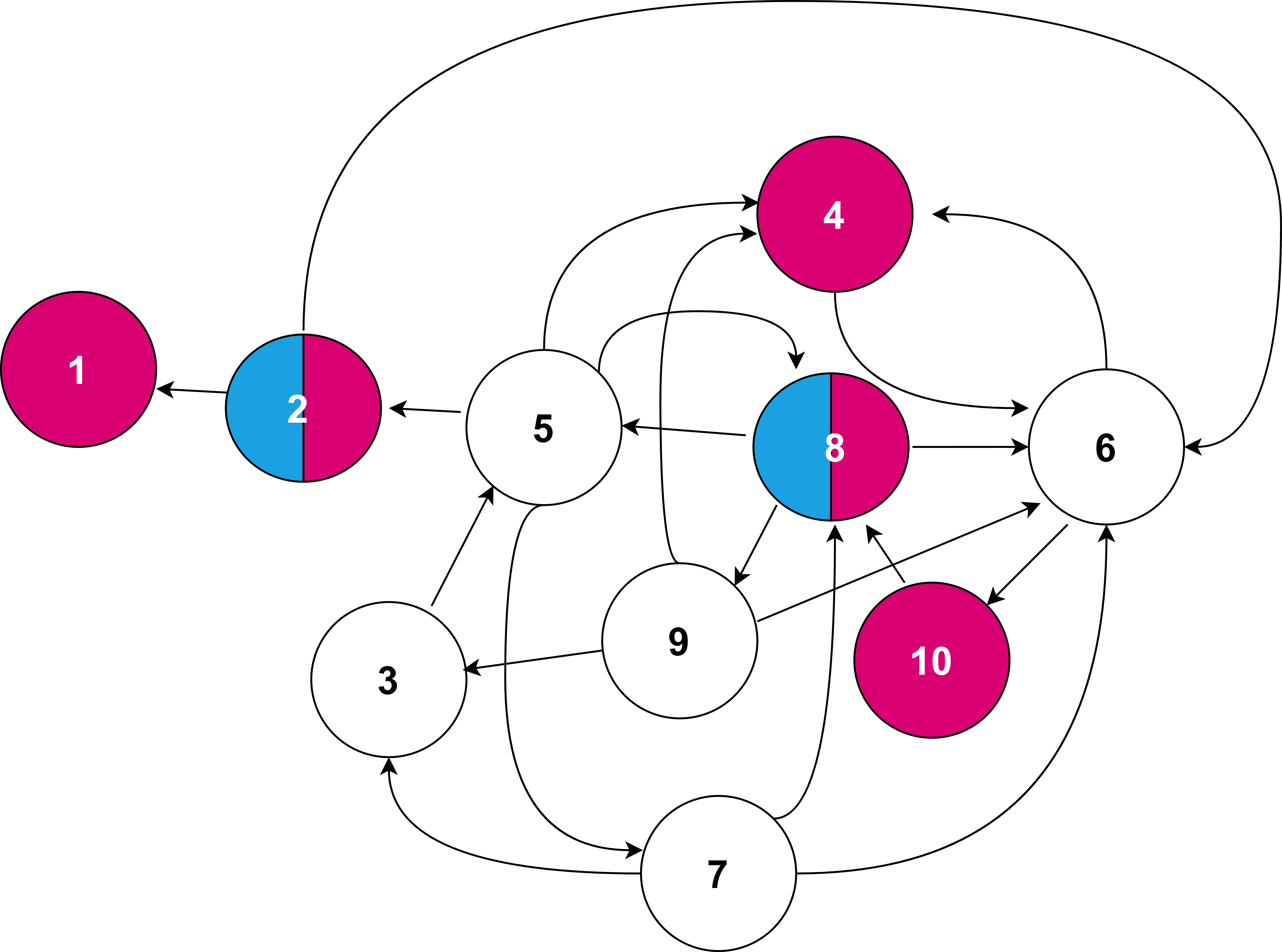}
\end{center}
\caption{Energy-optimal configuration of driver/target nodes in a random network with $N=10$ nodes and average degree $\langle k \rangle=2$, with fixed driver nodes at nodes $\{2,8\}$. Driver nodes are represented by cyan (grey in grayscale), and target nodes are magenta (dark grey). } 
\label{fig:example network}
\end{figure}

Next, we repeat the numerical experiment $1000$ times for target controlling $5$ nodes in a $N=10$ random network (ER network), with average degree $\langle k \rangle=2$, as shown in Fig. \ref{fig:example network}. The driver nodes position are fixed, and generically ensures full controllability of the network, regardless of the choice of $C$. The global energy-optimal configuration of target nodes placement is $\{1,2,4,8,10\}$ and TPGM was able to find this configuration $7.1 \%$ of the time. However, TPGM was also able to find the next-best performing configuration, which yields energy cost of the same order of magnitude, $\{2,4,6,8,5 \}$, with success rate of $13.9 \%$. Thus, generally, it can be seen that when path distances are reduced, control energy is minimized.

\begin{figure}[h!] 
\begin{center}
\includegraphics[scale=0.06]{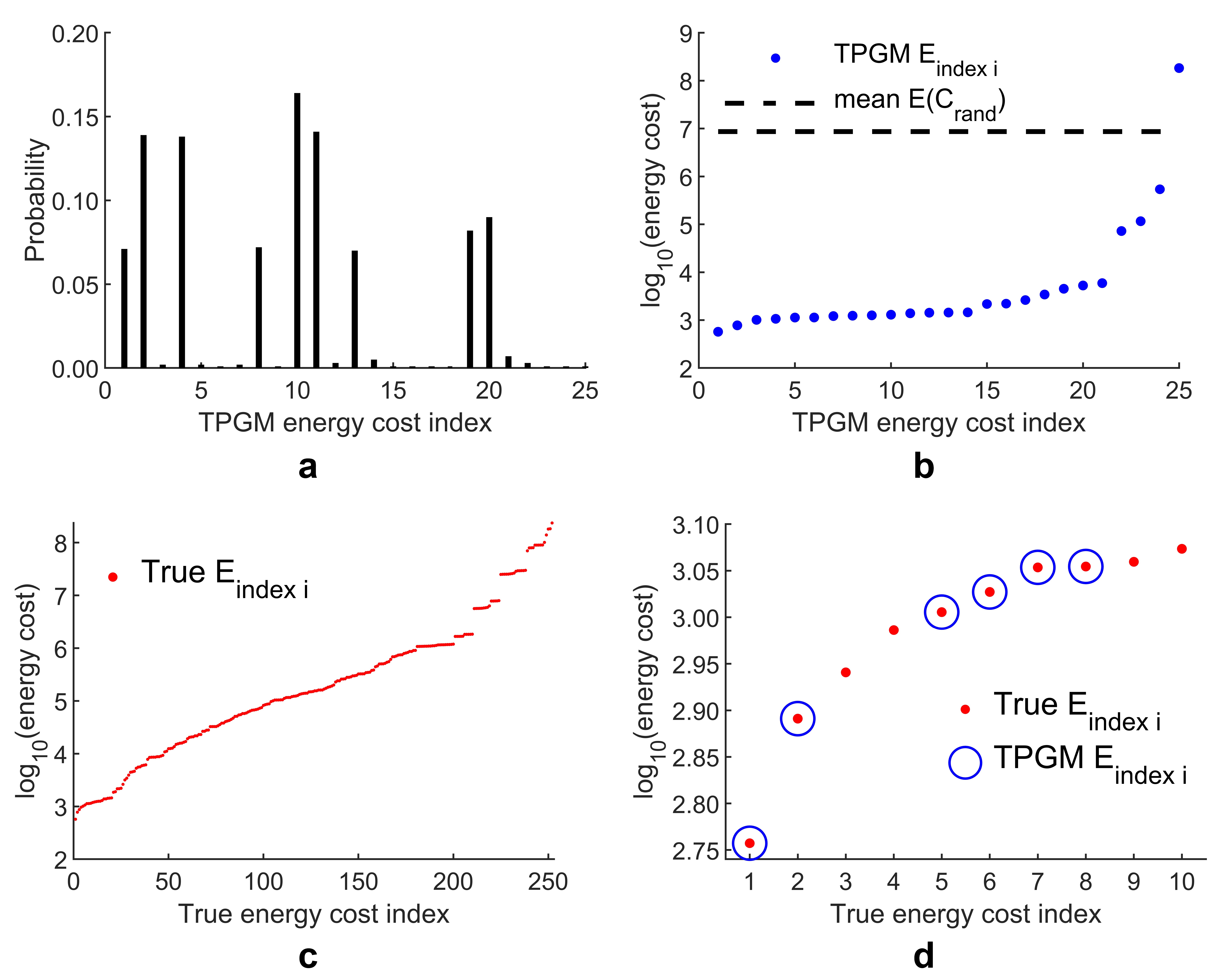}
\end{center}
\caption{Performance of TPGM for small random network. {\bf (a)} Plot of the probability of finding a specific target node set corresponding to each TPGM energy cost index $i$, TPGM $E_{\text{index i}}$, given in ascending order of energy cost. {\bf (b)} Plot of associated energy costs to each TPGM energy cost index. {\bf (c)} Plot of all $252$ energy costs in log scale for each named true $E_{\text{index i}}$, given in ascending order of energy cost. {\bf (d)} Plot of the 10 lowest true energy cost as shown by the red dots, of which $6$ are found by TPGM, represented by blue circles.  } 
\label{fig:example network performance}
\end{figure}

We quantify the performance of TPGM on the small random network in Fig.~\ref{fig:example network performance} In Fig.~\ref{fig:example network performance}(a), we display the probability mass function of TPGM obtaining all the solutions found, $C^*_{\text{binary}}$, and their associated energy costs, indexed in ascending order in energy, TPGM $E_{\text{index i}}$. It should be noted that all $25$ TPGM energy cost indexes are found from $1000$ independent iterative searches, and are not exhaustive. 
Fig.~\ref{fig:example network performance}(b) shows the energy costs associated to each TPGM $E_{\text{index i}}$. The full range of all possible energy costs is $^{10}C_5  = 252$, so 25 is roughly the top 10\% best solutions. For comparison, the average energy cost of selecting $5$ target nodes randomly without repetition, $\la E(C_{\text{rand}}) \ra$, is also plotted and represented by the black dashed lined on the same graph.
In Fig.~\ref{fig:example network performance}(c), we show all $252$ energy costs found from a brute force search. Matching the solutions found from TPGM to the true list of all possible energy costs, we find that the top $6$ solutions from TPGM belong to the true top $10$ energy costs, as shown in Fig.~\ref{fig:example network performance}(d). Therefore, while the rate of TPGM finding suboptimal solutions is high, the energy costs performances of the suboptimal solutions are in general comparable to the true optimal energy cost, $E_{\text{index 1}}$. From inspection, we see that the top 15 solutions have very similar energy, which may explain why TPGM only found the global solution (index 1) $7.1 \%$ of the time. In addition, the solutions found from the search algorithm also tend to lie within the neighborhood of the lower end of all possible energy costs. Besides, they also outperform the random selection scheme by at least a few orders of magnitude.

\subsection{Simulation results on complex networks}
We apply the TPGM to different complex networks such as random networks (ER network), scale-free networks (SF network), as well as various real networks spanning a diverse range: electronic circuit networks, food web networks, and social networks. For each network, we perform the iterative search 10 times and compare the performance of the obtained solution $C^{*}$ as well as the associate binary optimal solution $C^{*}_{\text{binary}}$ and compare them to heuristic selection schemes such as random selection (repeated over $100$ independent realizations) and degree-based selection of target nodes. We select $40\%$ of the nodes to be driver nodes using the standard way \cite{liu2011controllability,hopcroft1973n} to ensure controllability, and pick the remaining nodes randomly. For each network, once the driver nodes are picked, they remain fixed.  

\begin{figure}[h!]
\begin{center}
\includegraphics[scale=0.065]{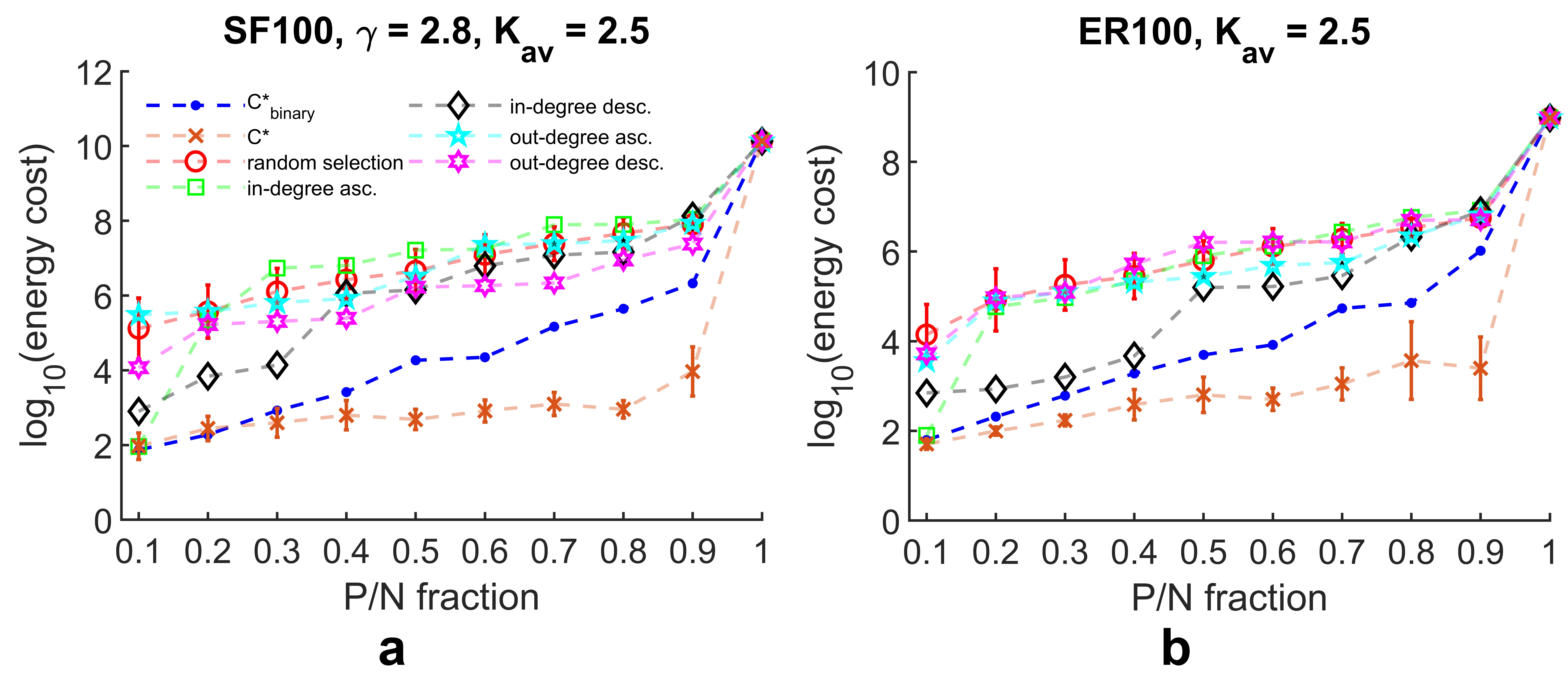}
\end{center}
\caption{Control energy needed for controlling target nodes of increasing cardinality. Optimal target node sets control energies, $\mathcal{E}(C^*)$ and $\mathcal{E}(C^*_{\text{binary}})$, are generally a few orders of magnitude better than selecting target node set from heuristic selection schemes. {\bf (a)} SF network and {\bf (b)} ER network.}
\label{model_networks}
\end{figure}

For the model networks of N=100 SF ($\gamma =2.8$) and ER, with average degrees $k_{av}=2.5$, we plot the control energy needed for controlling target node set of varying cardinality $P/N= \{0.1,0.2,...,0.9,1.0\}$ in Fig.~\ref{model_networks}(a) and (b). Consistent with the previous findings \cite{klickstein2017energy}, we find that the control energy scales exponentially with the cardinality of the target node set, regardless of the target node selection scheme chosen. However, when comparing the control energy of various selection strategies, our results show that generally, as compared to heuristic selection schemes, the energy cost for controlling target node set as found by TPGM, $C^*$ and $C^*_{\text{binary}}$, is lower by a few orders of magnitude. For the result of Fig.~\ref{model_networks}(b), the performance of the in-degree descending selection scheme (meaning that we choose $P/N \%$ of nodes that are ranked within the top $P/N \%$ of largest in-degrees) in the region $P/N= \{0.1,0.2,0.3,0.4\}$ is generally similar to optimal target node set $C^{*}_{\text{binary}}$. This is likely due to the fact that the driver nodes which were randomly selected happened to coincide with the nodes which have high in-degrees, which results in reduced path distances between driver nodes and target nodes, and thus reduced control energy cost. 

Finally, we apply TPGM to real networks and model networks of N=300, and compile the obtained results in table \ref{TPGM versus random}, comparing with the control energy of initial random selection $\mathcal{E}(C_0)$ and random selection $\mathcal{E}(C_\text{{rand}})$. With the exception of Circuit-s838 and Teacher-student, where we drive the network with $M/N=0.5$ and $M/N=0.6$ fractional number of driver nodes to target control $P/N=0.7$ and $P/N=0.75$ fractional number of nodes, each of the network is driven by $M/N=0.4$ fractional number of driver nodes to target control $P/N=0.6$ fractional number of nodes. The energy cost of target node set selected by degree-based selection scheme is presented in Table~\ref{degbased selection}. 

The heuristic selection schemes lead to target node sets $C_{\text{rand}}$, $C_{\text{in.asc}}$, $C_{\text{in.dsc}}$, $C_{\text{out.asc}}$, $C_{\text{out.dsc}}$. Respectively, $C_{\text{rand}}$ corresponds to random selection of target nodes, where repeats are disallowed; $C_{\text{in.asc}}$ refers to target nodes chosen in ascending order according to their weighted in-degrees. Thus, when choosing target nodes for $C_{\text{in.asc}}$, nodes with the lowest weighted in-degrees are selected. Likewise, $C_{\text{in.dsc}}$ is associated with target nodes chosen in descending order of weighted in-degrees, and nodes with the largest weighted in-degrees are selected. Analogously, $C_{\text{out.asc}}$ relates to target nodes picked in ascending order of weighted out-degrees; and $C_{\text{out.desc}}$ identifies target nodes of descending order of weighted out-degrees.

\begin{table}[h!]
%\scriptsize
%\footnotesize
 \begin{center}
   \caption{Control energy needed in various networks when different strategies are applied to selecting target control matrix $C$. $\langle\mathcal{E}(C_0)\rangle$ is the average control energy from $10$ independent initializations of random matrix without optimization. $\langle \mathcal{E}(C^*) \rangle$ is the average control energy from 10 independent iterative search using TPGM.  We then convert these ten solutions into binary matrices $\mathcal{E}([C^*_{\text{binary}}]^t)$, for $t={1,2,...,10}$.
The lowest of these is chosen as $\mathcal{E}(C^*_{\text{binary}})$.
 $\langle \mathcal{E}(C_{\text{rand}})\rangle$ is the average control energy of $100$ independent realizations of selecting target nodes randomly. Additional information such as standard deviations and mean of $\mathcal{E}([C^*_{\text{binary}}]^t)$ is presented in Supplementary Information. } 
   \label{TPGM versus random}
   \begin{tabular}{ccccccc}
   \\ [-2ex]
   \hline \\ [-2ex]
    Network & $N$ & edges & $\langle\mathcal{E}(C_0)\rangle$ & $\langle \mathcal{E}(C^*) \rangle$ &  $\mathcal{E}(C^{*}_{\text{binary}})$ & $\langle \mathcal{E}(C_{\text{rand}})\rangle$ \\\\ [-2ex]
    \hline
    \\ [-2ex]
    \multicolumn{1}{c}{\textbf{Model}}\\
    \hline
    \\[-2ex]
    SF300 & $300$ & $750$ & $8.11$E$07$ & $2.88$E$03$ & $2.65$E$05$ & $1.48$E$08$  \\
    ER300 & $300$ & $750$ & $1.88$E$06$ & $5.34$E$02$ & $1.24$E$05$ & $2.27$E$06$ \\
    \hline
    \\[-2ex]
    \multicolumn{1}{c}{\textbf{Electronic circuit }\cite{milo2004superfamilies}}\\
    \hline
    \\[-2ex]
    Circuit-s838 & $512$ & $819$ & $6.02$E$05$ & $4.48$E$02$ &$1.97$E$04$&$1.45$E$06$\\
    Circuit-s420 & $252$ & $399$&$6.01$E$04$&$1.33$E$02$&$6.92$E$03$&$8.97$E$05$\\
    Circuit-s208 & $122$ & $189$&$2.34$E$04$&$6.90$E$01$&$1.99$E$03$&$1.74$E$08$ \\
    \hline
    \\[-2ex]
    \multicolumn{1}{c}{\textbf{Food web} \cite{baird1998assessment,almunia1999benthic ,correll}}\\
    \hline
    \\[-2ex]
    StMarks & $54$ & $356$ &$4.83$E$03$&$6.31$E$01$&$2.77$E$02$&$5.54$E$03$ \\
    Maspalomas & $24$ & $82$&$1.76$E$03$&$3.72$E$01$&$3.65$E$01$&$2.59$E$04$ \\
    Rhode & $19$ & $53$&$8.07$E$02$&$2.00$E$01$&$3.65$E$01$&$1.46$E$05$\\
    \\[-2ex]
      \hline
      \\[-2ex]
    \multicolumn{1}{c}{\textbf{Social Influence} \cite{burt1987social,white1989rethinking}}\\
    \\[-2ex]
    \hline
    \\[-2ex]
    Phys-discuss-rev &$231$&$565$ &$1.90$E$04$&$1.99$E$02$&$4.11$E$03$&$1.49$E$05$ \\
    Teacher-student &$60$&$94$&$1.41$E$02$&$3.96$E$01$&$6.60$E$01$&$1.78$E$02$\\
    \\[-2ex]
     \hline
    \\[-2ex]
    \multicolumn{1}{c}{\textbf{Social} \cite{burt1987social,coleman1964introduction}}\\
    \\[-2ex]
    \hline
    \\[-2ex]
    Phys-friend-rev &$228$&$506$&$2.61$E$04$&$2.46$E$02$&$3.59$E$03$&$2.55$E$04$\\
    Highschool &$70$&$366$&$4.32$E$04$&$1.91$E$02$&$5.45$E$02$&$4.06$E$04$\\
    \\[-2ex]
     \hline
    \\[-2ex]
   
   \end{tabular}
 
 \end{center}
\end{table}

\begin{table}[h!]
%\scriptsize
%\footnotesize
 \begin{center}
 
   \caption{ Control energy needed in various networks when degree-based selection strategies are applied to selecting target control matrix $C$. $\mathcal{E}(C_{\text{in.asc}})$ refers to choosing target nodes in ascending order according to their weighted in-degrees. $\mathcal{E}(C_{in.dsc})$, $\mathcal{E}(C_{\text{out.asc}})$, and $\mathcal{E}(C_{\text{out.dsc}})$ follows analogously, where dsc refers to order of descending and out refers to weighted out-degree.} 
   \label{degbased selection}
   \begin{tabular}{ccccccc}
   \\ [-2ex]
   \hline \\ [-2ex]
    Network & $N$ & edges & $\mathcal{E}(C_{\text{in.asc}})$ & $\mathcal{E}(C_{\text{in.dsc}})$ &  $\mathcal{E}(C_{\text{out.asc}})$ & $\mathcal{E}(C_{\text{out.dsc}})$ \\\\ [-2ex]
    \hline
    \\ [-2ex]
    \multicolumn{1}{c}{\textbf{Model}}\\
    \hline
    \\[-2ex]
    SF300 & $300$ & $750$ & $2.99$E$08$ & $1.35$E$07$&$2.29$E$08$&$5.23$E$07$ \\
    ER300 & $300$ & $750$ & $4.18$E$06$&$9.18$E$04$&$6.51$E$05$&$2.39$E$06$ \\
    \hline
    \\[-2ex]
    \multicolumn{1}{c}{\textbf{Electronic circuit}}\\
    \hline
    \\[-2ex]
    Circuit-s838 & $512$ & $819$ & $4.77$E$05$&$4.99$E$05$&$4.06$E$04$&$4.80$E$05$\\
    Circuit-s420 & $252$ & $399$&$8.67$E$04$&$6.68$E$04$&$8.78$E$05$&$9.59$E$04$\\
    Circuit-s208 & $122$ & $189$&$3.20$E$06$&$2.65$E$04$&$4.00$E$04$&$3.20$E$06$ \\
    \hline
    \\[-2ex]
    \multicolumn{1}{c}{\textbf{Food web}}\\
    \hline
    \\[-2ex]
    StMarks & $54$ & $356$ &$4.73$E$02$&$4.53$E$02$&$7.16$E$02$&$5.19$E$02$ \\
    Maspalomas & $24$ & $82$&$7.52$E$02$&$1.34$E$03$&$1.04$E$03$&$9.37$E$02$ \\
    Rhode & $19$ & $53$&$6.52$E$07$&$1.35$E$06$&$6.11$E$07$&$5.94$E$06$\\
    \\[-2ex]
      \hline
      \\[-2ex]
    \multicolumn{1}{c}{\textbf{Social Influence}}\\
    \\[-2ex]
    \hline
    \\[-2ex]
    Phys-discuss-rev &$231$&$565$ &$4.29$E$04$&$2.62$E$04$&$1.11$E$04$&$2.95$E$05$\\
    Teacher-student &$60$&$94$&$8.89$E$01$&$7.21$E$02$&$4.14$E$02$&$9.41$E$01$\\
    \\[-2ex]
     \hline
    \\[-2ex]
    \multicolumn{1}{c}{\textbf{Social}}\\
    \\[-2ex]
    \hline
    \\[-2ex]
    Phys-friend-rev &$228$&$506$&$6.57$E$03$&$1.02$E$05$&$7.49$E$03$&$6.99$E$04$\\
    Highschool &$70$&$366$&$9.45$E$04$&$1.55$E$03$&$4.10$E$03$&$1.55$E$05$\\
    \\[-2ex]
     \hline
    \\[-2ex]

   \end{tabular}
 
 \end{center}
\end{table}

Examining table \ref{TPGM versus random}, we observe that for any network, starting from its initial, $\mathcal{E}(C_0)$, TPGM algorithm iteratively updates the $C_k$ matrix in the direction of quickest decreasing control energy until we obtain the convergent solution, $\mathcal{E}(C^*)$, where the energy cost is typically reduced by a few orders of magnitude. Based on section~\ref{select from real to discrete}, we can choose the target nodes from the obtained optimal solution, $C^*$, to obtain $C^*_{\text{binary}}$. While the conversion from dense real matrix solution to sparse binary matrix will in general result in increased control energy, comparing the energy cost $\mathcal{E}(C^*_{\text{binary}})$ to heuristic selection strategies of target nodes, such as $\mathcal{E}(C_{\text{rand}})$ in table \ref{TPGM versus random} and degree-based selection in table \ref{degbased selection}, we observe that the control energy compares favourably.

\section{Discussion} \label{Discussion}
\begin{figure}[h!]
\begin{center}
\includegraphics[scale=0.06]{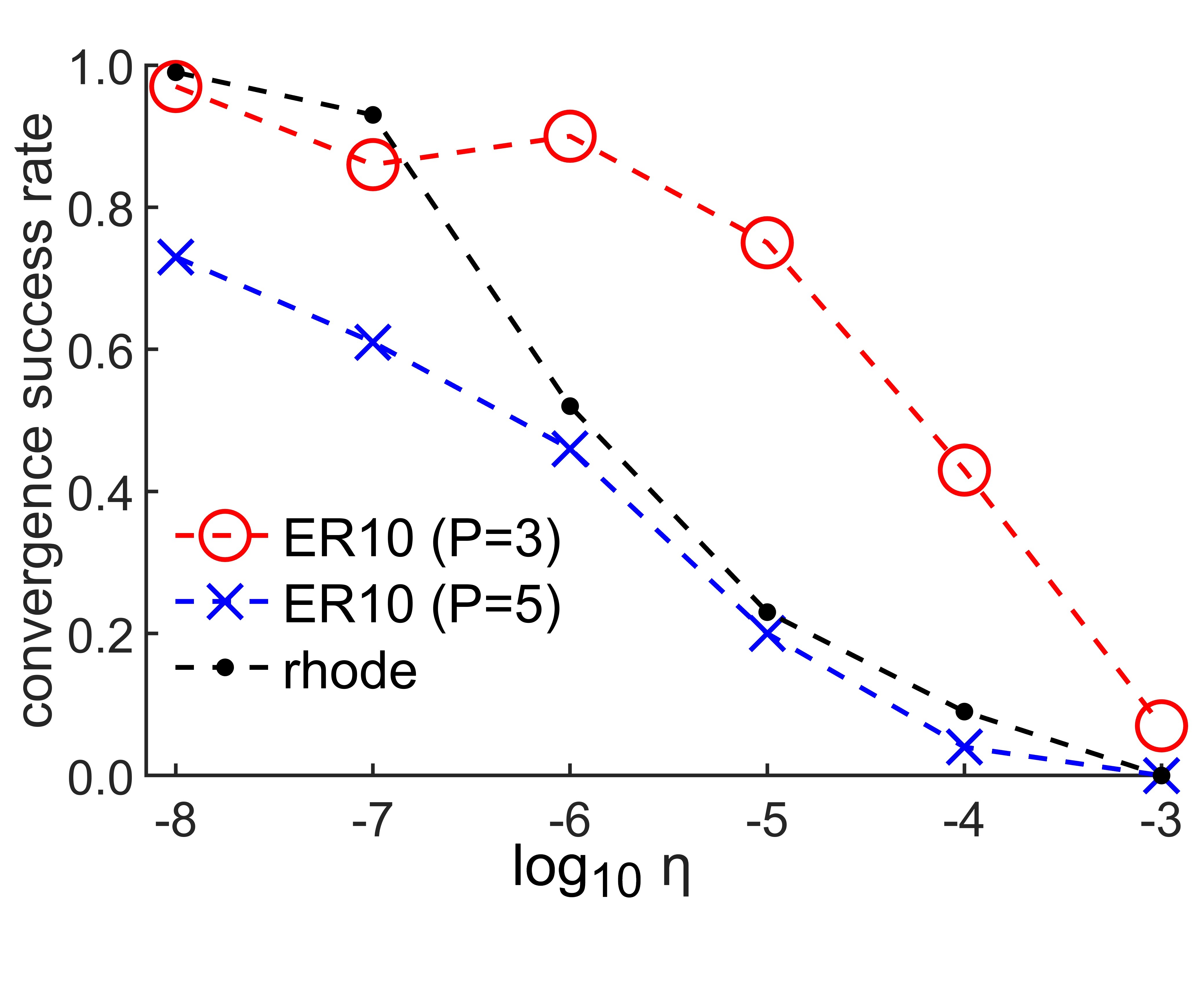}
\end{center}
\caption{The success rates of convergence as a function of $\log_{10} \eta$ for ER10 network (of Fig.~$2$) for target controlling $3$ and $5$ nodes, and rhode food web network (of table $1$) for target controlling $P/N=0.6$ number of nodes. $\eta$ varies from $1\text{e-}8$ to $1\text{e-}3$. Each data point is calculated based on $100$ independent iterative searches. } 
\label{fig:convergence_rate_versus_eta}
\end{figure}

When the learning rate $\eta$ is chosen appropriately, TPGM is convergent. To be illustrative, the success rates of convergence as a function of $\eta$ for different network topologies are shown in Fig.~\ref{fig:convergence_rate_versus_eta} Generally, when $\eta$ is lower, the success rate is higher. Furthermore, when $P$ is lower and the search space is smaller, higher $\eta$ tend to be accommodated. All $C^*$ results presented in this work are convergent, even though the chosen $\eta$ parameter may not have unity success rate. This is because, programmatically, for each independent iterative search, we can check the number of steps that the algorithm has taken; if the number of steps is abnormally small, the iteration is deemed to be non-convergent and to have terminated prematurely due to computation errors. Accordingly, we repeat the search iteration and in this way, convergence is always guaranteed.

\begin{figure}[h!]
\begin{center}
\includegraphics[scale=0.06]{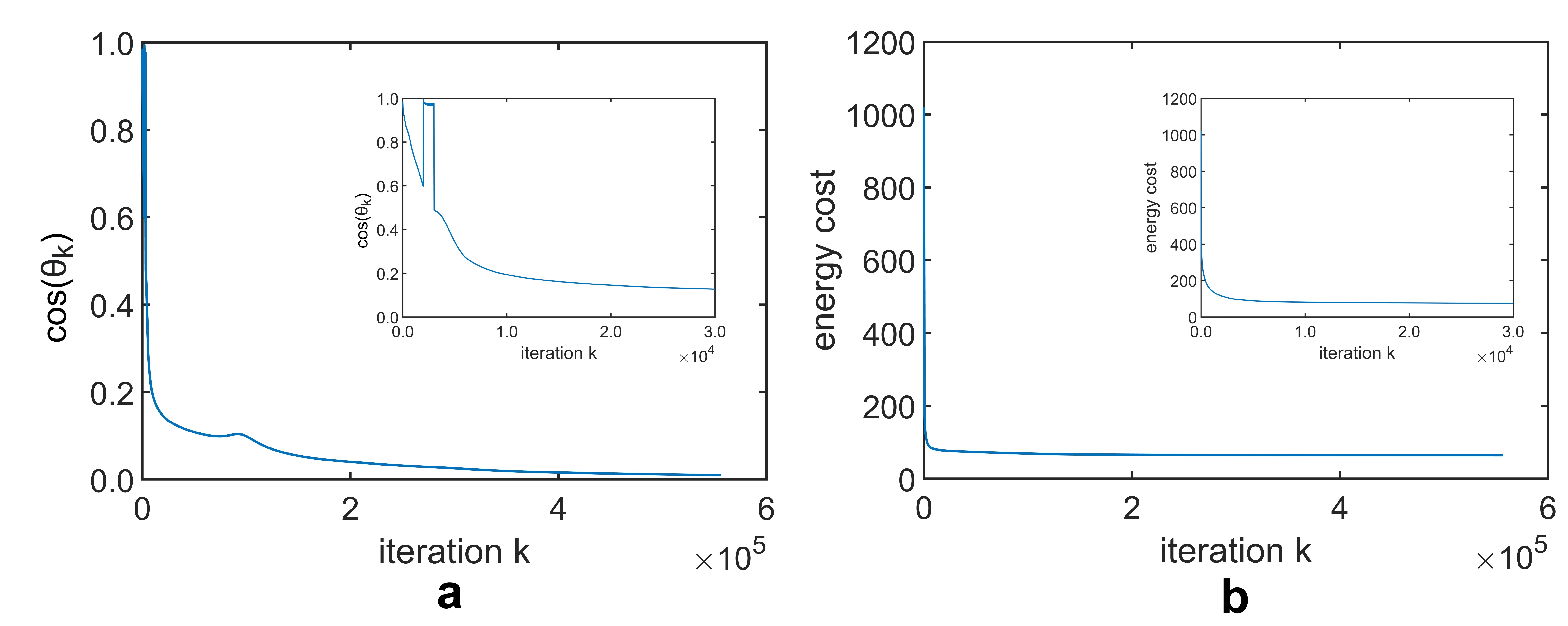}
\end{center}
\caption{ An illustration of one particular iterative search's convergence process on the electronic circuit network, Circuit-s208. {\bf (a)} $\cos(\theta_k)$ moves closer to zero with iteration, although it is not always non-increasing. {\bf (b)} Control energy is always non-increasing with each iteration, moving in the direction of largest decreasing energy cost. Insets show the first $0.3\times 10^5$ iterations.}
\label{convergence_withinset}
\end{figure}

The convergence process is shown in Fig.~\ref{convergence_withinset}(a) and (b). In Fig.~\ref{convergence_withinset}(a), the $k$-th step cosine angle, $\cos(\theta_k)$, between energy gradient $\nabla \mathcal{E}(C^T_k)$ and projected energy gradient $\mathcal{T}_{\tilde{C}^T_k} \nabla \mathcal{E}(C^T_k)$, approaches zero as TPGM iteration increases. The algorithm is deemed to have converged when the gradients are perpendicular. In Fig.~\ref{convergence_withinset}(b), we observe that each iteration of TPGM brings the matrix $C_k$ towards a lower energy cost, starting with a steep decrease in energy cost which becomes less steep with each iteration, until convergence.

It is important to initialize $C^T_0$ properly. In our experimentations, all elements of $C^T_0$ are chosen randomly and set to be $C^T_{ij}=1$ if node $i$ is chosen to be the $j\text{-th}$ target node initially, otherwise all other elements are set to be zero. We ensured row/ column linear independence, which necessitates that no nodes are chosen repeatedly. If row/ column linear independence is not adhered, computations would not proceed due to mathematical error. While TPGM can accommodate initial $C^T_0$ to be a random matrix with dense entries of random variables, this is ill-advised as the computation starting from a dense matrix is in general inefficient. Furthermore, the solutions $C^*$ and its associated $C^*_{\text{binary}}$ also tend to perform poorly when compared to the former approach. However, no matter the initial condition, convergence is only affected by the learning rate parameter $\eta$. 

When the number of nodes are small, such as in those shown in Fig.~\ref{fig:elementary topologies} and \ref{fig:example network}, we can systematically verify TPGM's searched solutions and compare them with the true optimal target node set, which can be found via a brute force exhaustive search approach. As $N$ increases, the $N \text{ choose }P$ brute force search is no longer viable. However, such limitation should also be viewed as a strength, because it shows the difficulty of the search problem due to its infeasibly large search space; by relaxing matrix $C$ into continuous search space, TPGM is able to search for $C^*$ in the direction of quickest energy cost decrease. Correspondingly, $C^*_{\text{binary}}$ is recovered as the `best signal' from $C^*$.  

The computational complexity of TPGM is $\mathcal{O}(N^3)$ owing to its iterative matrix multiplication. It should be noted that the computations of the controllability Gramian matrix, $W=\int_{t_0}^{t_f} e^{A\tau}BB^{T}e^{A^{T}\tau} d \tau$ as well as the calling of the exponential function for computing $e^{At_f}$ ($e^{A^Tt_f}$) are costly, which can result in bottlenecks when calculating them in the loop each time. An efficient way to code the TPGM is to compute the controllability Gramian $W$ and $e^{At_f}$ ($e^{A^Tt_f}$) outside the while loop and storing them as variables to be retrieved within the loop. Furthermore, instead of directly computing the integral, an efficient way to compute the controllability Gramian is to use the method of ref. \cite{van1978computing}.

The condition number of the controllability Gramian matrix plays an important role in our numerical experiments. When trying to target control a network using only driver nodes found from structural controllability \cite{liu2011controllability}, the condition number of the Gramian may be too high and results in an infeasibly high control energy \cite{wang2017physical}. Increasing the number of control signals can lower the condition number and render the computation of the Gramian feasible \cite{sun2013controllability,wang2017physical}. Note that the condition number is dependent on matrices $A$, $B$, as well as time horizon $t_f$.  For some networks, despite increasing the number of driver nodes, the condition number of the Gramian may still be infeasibly high. To lower the condition number, we can normalise the link weights of the connection matrix, $\{a_{ij}\}$, by dividing throughout a normalization constant. For networks whose connection strengths are not specified, we set $\{a_{ij}\}=1$ if there is a directed link from node j to node i. However, if this results in an infeasible condition number, then we will set the link strength to be random uniform $[0.5,1.5]$, which tends to make the condition number feasible.

Building a connection between target and driver node set optimization could be very interesting. Drawing on driver nodes optimization of previous works \cite{li2018enabling,ding2017key,li2015minimum,li2016optimal,li2018optimization} for controlling the full node set, the conclusions formed are as follows: when driver nodes are placed in such a way such that they are evenly spaced out, the control energy is most optimal. Thus, by grouping controlled nodes to their nearest driver nodes and arranging the driver nodes in such a way that no groups have excessive geodesic paths \cite{newman2003structure} from drivers to controlled nodes, energy cost is most optimized.  We expect the same conclusion to hold even for the case of target control, because full control and target control are not fundamentally different. Further, we expect that for the reverse problem of optimizing $B$, given that $C$ is fixed, the same energy-optimal configurations, as presented in Fig.~\ref{fig:elementary topologies}
and Fig.~\ref{fig:example network}, to be found. In other words, whether optimizing driver nodes placement, or target nodes arrangement, the energy-optimal configurations are the same.  Therefore, energy cost minimization could be an important mechanism in explaining structural self-organized configurations of driver/ target nodes in a natural or man-made complex system.

In conclusion, this paper demonstrate the possibility of minimizing control energy of directed complex networks by optimizing the target node set $C$, given that network connection matrix $A$, driver node placement $B$, and time horizon $t_f$ are fixed. We achieve this by adapting target control \cite{klickstein2017energy}, as well as deriving the energy gradient $\frac{\partial \mathcal{E}(C^T_k)}{\partial C^T} = \nabla\mathcal{E}(C^T_k) $ into the TPGM algorithm \cite{li2018optimization}. By ascribing target nodes to their nearest driver nodes, a directed complex network driven by driver nodes can be decomposed into elementary topologies \cite{liu2011controllability,li2018enabling}, such as stem, circle, and dilation. Through extensive simulations on elementary topologies, our results reveal that control energy is most optimal when target nodes are chosen such that path distances from driver nodes to target nodes are minimized, corroborating existing literature results \cite{chen2016energy,klickstein2018control}. Furthermore, we validate our results on model networks (ER and SF) and real networks and show that optimal target node set (both $C^*$ and $C^*_{\text{binary}}$) has control energy of a few orders of magnitude lower compared to target nodes chosen from heuristic selection schemes, such as random or nodes degree-based selection. Compared to previous works of control energy optimization \cite{li2015minimum,gao2018towards,li2016optimal,li2018optimization,lindmark2018minimum}, which focus on driver node set optimization, we concentrate on target node set optimization, and show that in the context of target control, target node set can account for variability of the control energy of a few orders of magnitude. 

The problem of optimizing target node set in the interest of control energy could be applicable to linear opinion networks \cite{tanner2004controllability,liu2008controllability,rahmani2009controllability}, where the state vector represents opinions of individuals, and the driver nodes are modeled as agents of influence. In such a system, we may be interested in influencing a certain fractional share of opinions to align with a pre-defined favorable opinion. Much like a voter model problem \cite{masuda2015opinion}, where we are only interested in obtaining the majority share of opinions, the specificity of which individuals to target control is not so much important as the control energy, which we would like to minimize. There are some avenues of research which appear to be promising. For example, within the framework of network controllability, recent works have incorporated conformity behavior \cite{wang2015controlling,nie2018control}, where individuals' opinions adapt over time to mirror the average of their neighbors', thus making the network dynamics richer and more realistic. It would be interesting to explore how optimal target nodes relate to a linear opinion network, both with and without conformity behavior in the future.

\section{Methods}
{\bf Model networks.} Similar to recent works \cite{klickstein2017energy,yan2015spectrum}, the model networks considered in this paper are modeled with stable dynamics, i.e. $\{a_{ii}\} < 0$ $\forall i$. Specifically, we choose the diagonal of the connection matrix $A$ to be chosen random uniformly from $[-1,1]$, and then stabilize the nodal dynamics with  $\{a_{ii}\}=\delta_i + \epsilon$, where $\epsilon$ is chosen such that the eigenvalues of $A$ are all negative and the largest eigenvalue is $-1$. The scale-free model network is constructed from the static model \cite{catanzaro2005analytic,goh2001universal}, and the link weights $\{a_{ij}\}$ are drawn randomly from a uniform  interval $[0.5,1.5]$.

{\bf Input nodes.} The driver nodes selected are chosen using the method detailed in ref. \cite{liu2011controllability}, which uses the Hopcroft-Karp algorithm \cite{hopcroft1973n} to find the driver nodes to ensure controllability. To be specific, the driver nodes found in the research work presented in this paper are an overestimation, and guarantee full controllability of the complex network. Thus, controllability is always ensured, regardless of the choice of target node set $C$, and the term $(CWC^T)$ is always invertible.

{\bf Weighted nodes degree.} When considering nodes degree in the degree-based selection strategies, we computed the weighted link weights of each node to determine the in-degree (out-degree). For example, the in-degree of node i is computed as: $=\sum^N_{j=1,j\neq i} \{a_{ij}\}$. Note that self-links, $\{a_{ii}\}$, do not count as node degree as they arise from categorically different sources \cite{cowan2012nodal}. 

{\bf Gramian computation.} The controllability Gramian can be efficiently calculated using the method of ref. \cite{van1978computing}:

\begin{equation} \label{block}
\begin{aligned}
&\text{exp}
\begin{pmatrix}
\begin{bmatrix}
-A & BB^T \\
0 & A^T
\end{bmatrix}
t_f
\end{pmatrix}
=
\begin{bmatrix}
\text{F}_2(t_f) & \text{G}_2(t_f)\\
0 & \text{F}_3(t_f)
\end{bmatrix}
\end{aligned}
\end{equation}

\begin{equation} \label{gramian van}
\begin{aligned}
& W= \text{F}_3(t_f)^T \: \text{G}_2(t_f)
\end{aligned}
\end{equation}

\noindent Where each block partitioned matrix of (\ref{block}) is a $N \times N$ matrix. The controllability Gramian is computed using (\ref{gramian van}).

%\bibliographystyle{unsrt}
% keep biblographystyle{} commented when using 'wlscirep' scirep package
\bibliography{ref}

\section{Acknowledgement}
We would like to thank Tan Yew Lee for discussion on computation speed and bottlenecks and Dr. Lim Yi Xian for guiding us to create high quality figures. H.C. and E.H.Y. acknowledge support from Nanyang Technological University, Singapore, under its Start Up Grant Scheme (04INS000175C230).

\section{Author contribution}
H.C. conceived the project and performed the calculations under the supervision of E.H.Y.; Both authors wrote, reviewed, and revised the manuscript. 

\section{Additional information}
{\bf Supplementary information} accompanies this paper at [to be placed]\\
{\bf Competing interests:} The authors declare that they have no competing interests.

\end{document}